\documentclass{iau} 
\usepackage{graphicx}

\title[Galaxy Outskirts \& Dark Matter] %% give here short title %%
{What can the outskirts of galaxies tell us about dark matter?}

\author[Chris Power]   %% give here short author list %%
{Chris Power}

\affiliation{International Centre for Radio Astronomy Research, The University of Western Australia, 35 Stirling Highway, Crawley, Western Australia 6009, Australia\\ E-mail: {\tt chris.power@icrar.org}}

\def\lesssim{\mathrel{\hbox{\rlap{\hbox{\lower4pt\hbox{$\sim$}}}\hbox{$<$}}}}
\def\gtrsim{\mathrel{\hbox{\rlap{\hbox{\lower4pt\hbox{$\sim$}}}\hbox{$>$}}}}

\pubyear{2016}
\volume{xxx}  %% insert here IAU Symposium No.
\jname{IAUS321: Formation and Evolution of Galaxy OutskirtsIAUS32}
\editors{Armando Gil de Paz, Johan Knapen, Janice Lee, eds.}
\begin{document}

\maketitle

\begin{abstract}
  Deep observations of galaxy outskirts reveal faint extended stellar components 
  (ESCs) of streams, shells, and halos, which are ghostly remnants of the tidal 
  disruption of satellite galaxies. We use cosmological galaxy formation simulations 
  in Cold Dark Matter (CDM) and Warm Dark Matter (WDM) models to explore how the 
  dark matter model influences the spatial, kinematic, and orbital properties of 
  ESCs. These reveal that the spherically averaged stellar mass density at large 
  galacto-centric radius can be depressed by up to a factor of $\sim$10 in WDM 
  models relative to the CDM model, reflecting the anticipated suppressed abundance 
  of satellite galaxies in WDM models. However, these differences are much smaller 
  in WDM models that are compatible with observational limits, and are comparable in 
  size to the system-to-system variation we find within the CDM model. This suggests 
  that it will be challenging to place limits on dark matter using only the 
  unresolved ESC.
  
\end{abstract}

\firstsection % if your document starts with a section,
\section{Motivation}
The defining prediction of the Cold Dark Matter (CDM) model of cosmological structure
formation is that dark matter haloes contain an abundance of lower mass
substructure haloes (hereafter subhalos; e.g. \cite[Klypin et al. 1999, Moore et al. 1999]{klypin.etal.1999b,moore.etal.1999}). Alternatives to the CDM model, such as Warm
Dark Matter (WDM), suppress the abundance of low-mass dark matter halos, and
consequently the abundance of subhalos, but distinguishing between these alternatives
and CDM in this way in a robust fashion is challenging 
(e.g. \cite[Power 2013]{power.2013}). This is because the mass scale at
which differences between plausible dark matter models is most likely to be evident --
at or below the scale of the satellites of the Milky Way
(e.g. \cite[Schneider et al. 2014]{schneider.etal.2014}) -- is also the mass scale
at or below which galaxy formation is inefficient and apparently stochastic
(e.g. \cite[Boylan-Kolchin et al. 2011]{boylan-kolchin.etal.2011}).

However, deep observations of the Milky Way and external galaxies reveal that
they are embedded in a diffuse, extended stellar component (ESC) of shells, streams, and
halos (e.g. \cite[Freeman \& Bland-Hawthorn 2002, Helmi 2008]{freeman.bland-hawthorn.200
  ,helmi.2008}), which are the relics of the tidal disruption of satellite galaxies
(e.g. \cite[Bullock \& Johnston 2005, Cooper et al. 2010]{bullock.johnston.2005,cooper.etal.2010}). Because these satellites are embedded
within low-mass dark matter subhalos, it is plausible that the underlying dark matter
model could leave an imprint on the structure of the ESC -- models in which
the abundance of low-mass subhalos, and consequently satellite galaxies, is
suppressed could result in less massive, less luminous, and more centrally
concentrated ESCs compared to CDM. This is because subhalos follow
similar orbits in CDM and WDM models \cite[(e.g. Knebe et al. 2008)]{knebe.etal.2008}, and because
low-mass subhalos and satellites follow preferentially radial orbits and require many
pericentric passages before their orbits decay, their tidally stripped stars can be
spread over large galacto-centric distances. The physics that governs galaxy formation
should not depend on the underlying dark matter model, and so
differences in ESC properties should reflect differences in satellite abundance,
which depends on the subhalo abundance. These differences could be
accessible to current (\cite[e.g. van Dokkum et al. 2014]{dragonfly}) and future
surveys (\cite[e.g. LSST Science Collaboration 2009]{lsst}) that target the diffuse,
low surface brightness environs of galaxies, and so represents a potentially robust
observational test of dark matter. 
\section{Extended Stellar Components as an Observational Testbed for Dark Matter}
We have tested the feasibility of this idea in \cite{power.2016} using a set of cosmological zoom galaxy
formation simulations of 6 Milky Way mass galaxies (MW01 to MW06) forming in a CDM model, with
$M_{200} \simeq 2 \times 10^{12} h^{-1} {\rm M}_{\odot}$ and $R_{200} \simeq 200 h^{-1} {\rm kpc}$ at $z$=0. These were run
with a version of the {\small GADGET}
$N$-body/SPH code of \cite[Springel (2005)]{springel.2005} that includes prescriptions
for cooling, star formation, and supernova feedback.

In the case of one of the galaxies, MW02, we ran 3 additional versions, assuming WDM models with particle masses of $m_{\rm WDM}$=0.5, 1, and 2 keV/$c^2$; this
was done by modifying the initial CDM power spectrum following the approach of
\cite[Bode et al. (2001)]{bode.etal.2001}. Although models with
$m_{\rm WDM}<2 {\rm keV}/c^2$ are not favoured by current observational limits, we
wanted to establish whether or not dark matter models that alter the abundance of
substructure can leave an imprint on observable properties of galaxies.

\begin{figure}
  \centerline{\includegraphics[width=0.45\columnwidth]{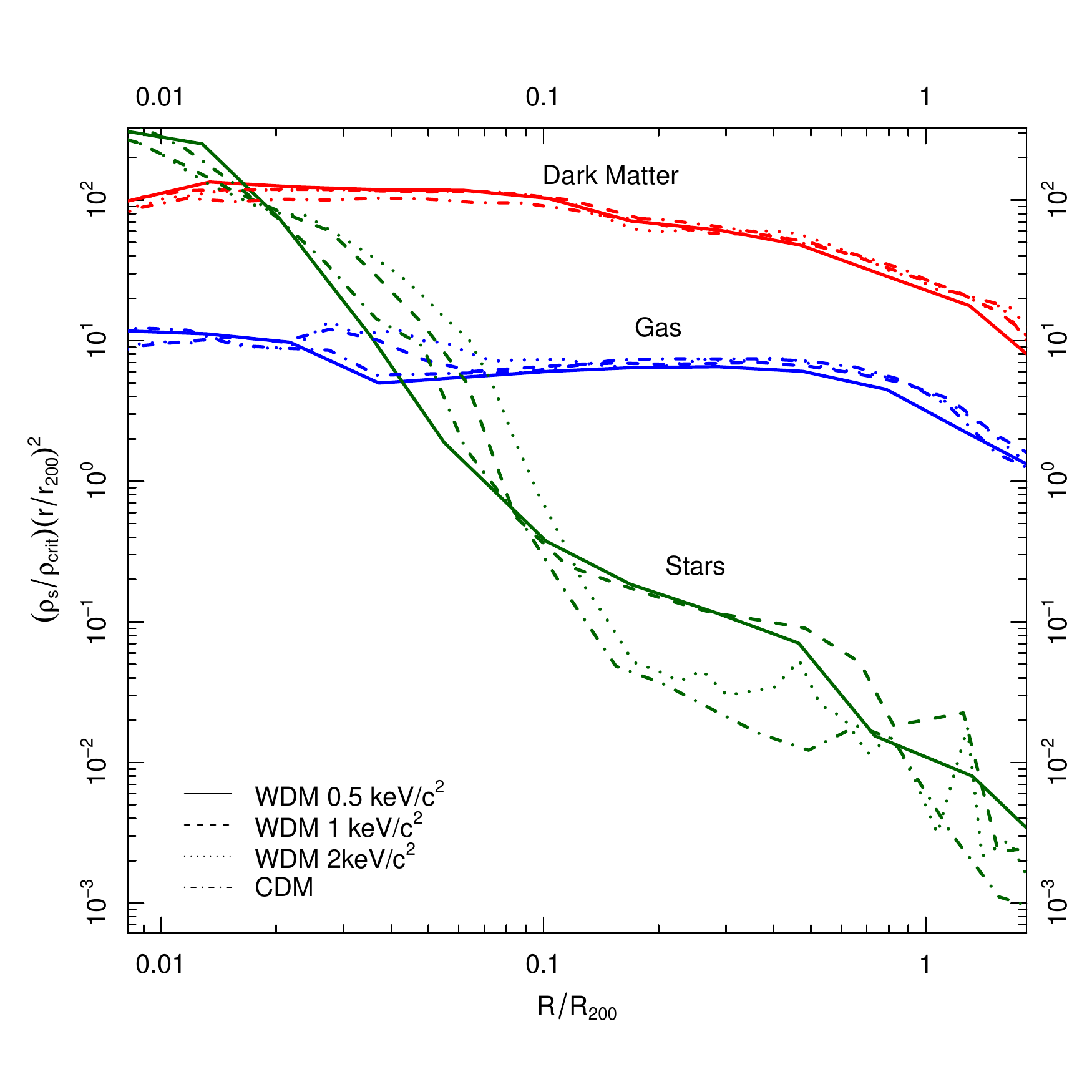}
    \includegraphics[width=0.45\columnwidth]{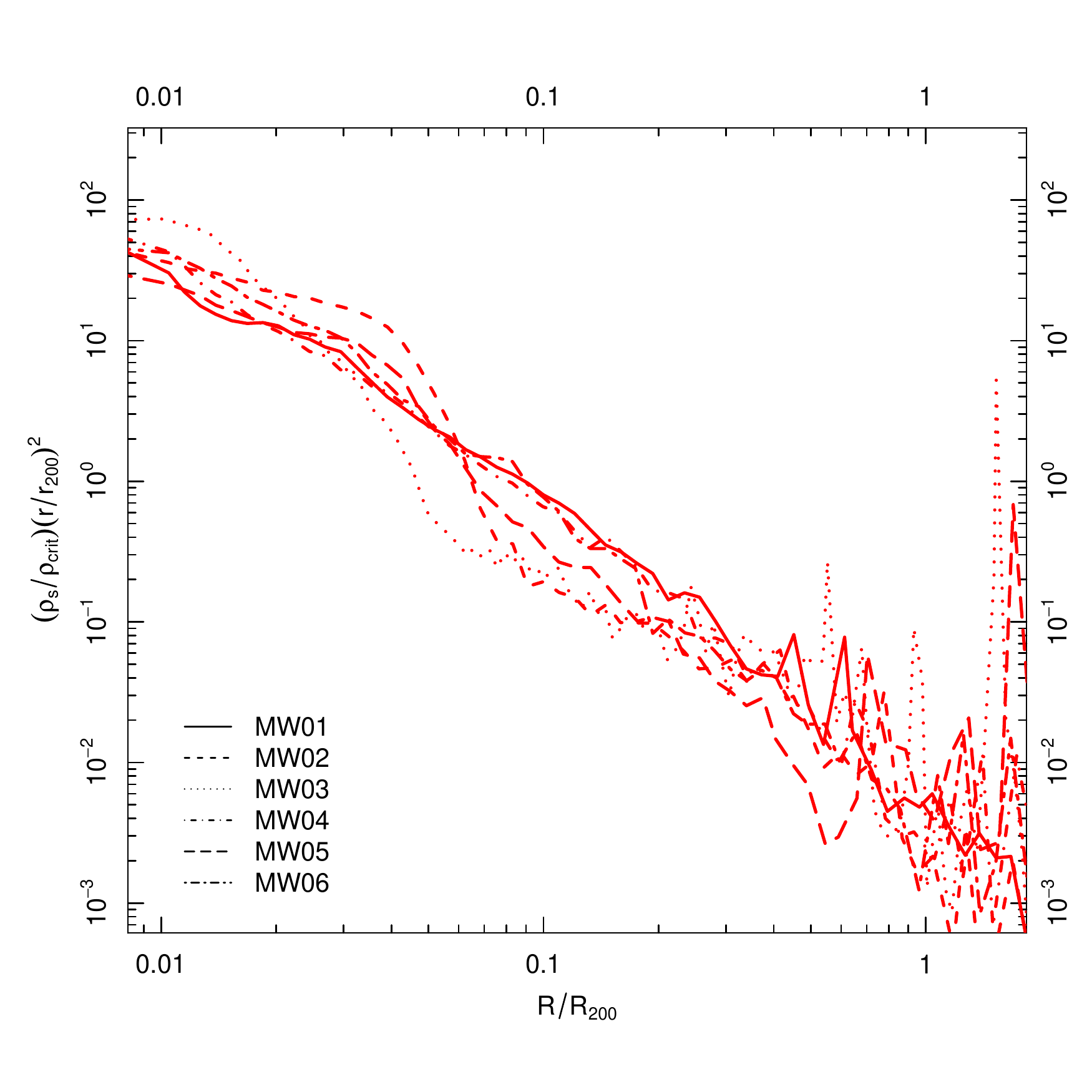}}
  \caption{{\bf Spherically Averaged Mass Profiles at $z$=0.} Here we show
    spherically averaged mass density profiles of different components of the simulated galaxies. On the \textit{left} we compare stars, dark matter, and gas inthe
    fiducial CDM run (dotted-dashed curves) and WDM $m_{\rm WDM}$=2, 1, and 0.5
    ${\rm keV}/c^2$ profiles (dotted, dashed, and solid curves).
    On the \textit{right} we compare only stellar profiles for the 6 galaxies that form
    in the CDM model, all selected to form in low-density regions and all with
    $M_{200}\simeq 2 \times 10^{12} h^{-1} {\rm M}_{\odot}$ at $z$=0.}
  \label{fig:density_profile_z0}
\end{figure}

These simulations reveal that it is in the spatial structure of the ESC
that the imprint of the underlying dark matter model is most readily apparent.
In the left hand panel of Figure~\ref{fig:density_profile_z0}, we show
spherically averaged mass density profiles at $z$=0 for stars, gas, and dark
matter for the the CDM and WDM $m_{\rm WDM}$=2, 1, and 0.5 keV/$c^2$ keV
versions of MW02 (dotted-dashed, dotted, dashed, solid curves),
while in the right hand panel we show only stellar profiles for the CDM
versions of the 6 galaxies, MW01 to MW06. This Figure confirms our expectation
that properties of the ESC varies with the underlying dark matter model --
outside of the region within which
the galaxy disc reside, between $R/R_{200}\sim 0.1$ and $R/R_{200}\sim 0.5$, we find
that the
spherically averaged stellar density is approximately an order of magnitude smaller in
the WDM $m_{\rm WDM}$=1 and 0.5 keV$/c^2$ runs than in the CDM and WDM 2 keV$/c^2$ runs.
This difference is imprinted early in the history of the galaxy at least back to
$z \simeq 4$ -- and persists to $z$=0. 

However, these significant differences are most readily apparent only in the more
exteme WDM models that we consider, with $m_{\rm WDM}<2 {\rm keV}/c^2$; for values of
$m_{\rm WDM}$ consistent with observational limits, differences with respect to the
favoured CDM model are comparable in size to the galaxy-to-galaxy variation in the
CDM model, as we can infer from the right panel of Figure~\ref{fig:density_profile_z0}.
This variation, which reflect differences in assembly history, highlights the
difficulty of using the ESC as a test of dark matter -- but suggests that the ESC can
be used to extend the concept of galactic archaeology to systems beyond the Milky Way
and Andromeda, as is being done in nearby galaxies
(\cite[e.g. Martinez-Delgado et al. 2010, van Dokkum et al. 2014]{martinez-delgado.etal.2010,dragonfly}), and
will become possible for statistical samples of galaxies within the Local Volume with, for example, {\small LSST} (\cite[LSST Science Collaboration 2009]{lsst}). Combining
metallicity, kinematics, and spatial structure, it should be possible to trace the
assembly history of galaxies.

\section{Prospects}
The focus of this work has been on the utility of the unresolved ESC -- the
spatial distribution and kinematics of stellar material in galaxy outskirts that
will be accessible to deep imaging surveys, possibly by stacking large
numbers of galaxies by central galaxy stellar mass or halo mass bins -- as a testbed
for dark matter, to carry out a statistical detection. Our results suggest that this
will be challenging. However, the resolved ESC might also be used, by combining
spatial, kinematic and metallicity substructure information to test dark matter;
this will require the kind of statistical sample of Local Volume galaxies that will be
accessible with {\small LSST}. Even if this remains a challenging test of dark
matter, there is good reason to expect that we can use properties of the ESC to
explore the mass assembly histories of galaxies, tracing merger and accretion events
using material in the outer halo, and placing limits
of the growth of galaxies in the context of their larger scale environment.

\end{document}